\documentclass[summary]{ursi}

\title{A model local interpretation routine for deep learning based radio galaxy classification}

\author{Hongming Tang*\affref{ref1}, 
        Shiyu Yue\affref{ref2},
        Zijun Wang\affref{ref2},
        Jizhe Lai\affref{ref3},
        Leyao Wei\affref{ref2},
        Yan Luo\affref{ref2},
        Chuni Liang\affref{ref2},
        Jiani Chu\affref{ref1}
        }

\affiliation{%
  \aff{ref1}{Department of Astronomy, Tsinghua University, Beijing 100084, China; e-mail: hongmingt@mail.tsinghua.edu.cn}
  \aff{ref2}{School of Physics and Astronomy, Sun Yat-sen University, 2 Daxue Road, Zhuhai 519082, China; e-mail: yueshy5@mail2.sysu.edu.cn}
  \aff{ref3}{School of Physics, Sun Yat-sen University, No. 135 Xingang Xi Road, Guangzhou 510275, P.R. China}
}

\begin{document}

\maketitle

\begin{abstract}
     Radio galaxy morphological classification is one of the critical steps when producing source catalogues for large-scale radio continuum surveys. While many recent studies attempted to classify source radio morphology from survey image data using deep learning algorithms (i.e., Convolutional Neural Networks), they concentrated on model robustness most time. It is unclear whether a model similarly makes predictions as radio astronomers did. In this work, we used Local Interpretable Model-agnostic Explanation (LIME), an state-of-the-art eXplainable Artificial Intelligence (XAI) technique to explain model prediction behaviour and thus examine the hypothesis in a proof-of-concept manner. In what follows, we describe how \textbf{LIME} generally works and early results about how it helped explain predictions of a radio galaxy classification model using this technique.\footnote{Hongming Tang and Shiyu Yue have equally contributed to this work.}
\end{abstract}

\section{Introduction}

Radio galaxy morphological classification is highly valued in radio astronomy as it reveals both the evolution process of a radio galaxy and how it interacted with the local environment \cite{Becker2021}. Motivated by the rapidly growing radio source sample number produced by large-scale radio continuum surveys (i.e., \cite{Norris2011, Norris2021}), people started to face the data challenge using machine learning.
In recent years, multiple deep learning algorithms have been developed to either find and classify radio galaxy morphology (i.e., \cite{Lao2021}) or do classification alone (i.e.,\cite{Becker2021, Bowles2021, Mohan2022}). Most of them have achieved human-comparable classification accuracy.

Besides the robust model performance these algorithms achieved, their model interpretability received less attention. Whether a deep learning algorithm is predicting radio galaxy morphology in the same way we radio astronomers did remains an ongoing question to answer. The latest effort to address this problem introduced a self-attention mechanism to their models, which enabled people to explain model prediction behaviour by looking at reasonably static image features from generated model attention maps \cite{Bowles2021}. However, when explaining many state-of-the-art radio galaxy classification algorithms, post-hoc model explanation methods that do not require model architecture manipulation would still be necessary.

Though radio galaxy classification system has become so complicated \cite{Rudnick2021}, Fanaroff and Riley binary classification (FR classification hereafter) system remains popular and used widely \cite{Fanaroff1974} since 1974. A radio galaxy would be identified as either edge-brightened sources (FR\,II)  or edge-darkened sources (FR\,I) \cite{Bowles2021}. In this work, we tried to explain a deep learning model developed for FR classification using an eXplainable Aritificial Intelligence technique called Local Interpretable Model-agnostic Explanation (LIME)\footnote{https://github.com/marcotcr/lime}. In order to perform model interpretation, we made used of FR-DEEP v2, a machine learning dataset for FR classification\footnote{https://github.com/HongmingTang060313/FRDEEP\_v2.0} to train a Convolutional Neural Network (CNN) based FR classification algorithm, reaching $\rm \sim 91 \%$ model general accuracy in a testset of 130 radio sources. Since we here focus on model interpretation, detailed model training and evaluation process would be shown in Tang and Yue et al. (2023, in prep.) instead. For the rest of this work, we would introduce the \textbf{Felzenszwalb} image segmentation method and how it combines with \textbf{LIME} in Section~\ref{sec:felzenszwalb} and ~\ref{sec:lime}. Section~\ref{sec:application} would shows our early results of how \textbf{LIME} help explaining model predictions, and we summarize our conclusion in Section~\ref{sec:conclusion}.

\section{Felzenszwalb}
\label{sec:felzenszwalb}

\textbf{Felzenszwalb} \cite{Felzenszwalb2004} is an graph-based image segmentation method. By seeing each image pixel as a \textbf{vertice}, neighbouring vertices are connected by \textbf{edges} along with weights measuring the dissimilarity of each vertice pair \cite{Felzenszwalb2004}. For any two neighbouring image segmented components C1 and C2, they can only stay independent to each other if their pairwise comparison \textbf{D(C1,C2)} satisfy:

\begin{equation}
  \label{eq:pairwise_comparison}
  \rm D(C1,C2) = True\,\,if\,\,Dif(C1,C2) > Mint(C1,C2)
\end{equation}

where \textbf{Dif(C1,C2)} denotes the minimum weight edge connecting C1 and C2. \textbf{Mint(C1,C2)} represents the minimum internal difference (\textbf{int(C)}; the largest weight in the neighbouring spanning tree of a component C) considering both component C1 and C2: 

\begin{equation}
  \label{eq:mint}
  \rm Mint(C1,C2) = Min(int(C1)+ \frac{k}{\lvert C1 \rvert}, int(C2)+ \frac{k}{\lvert C2 \rvert})
\end{equation}

where k is a constant to control the preference of having larger or smaller segments, and $\lvert C1 \rvert$ corresponds to the size of component C1 (similar for $\lvert C2 \rvert$). The two components will merge if D(C1, C2) equals \textbf{\it False}. 

Compared with other segmentation methods, \textbf{Felzenszwalb} can generate image segments or \textbf{super-pixels} neither not "too coarse" nor "too fine", making it an appropriate tool when objects of interest in an image share modest sizes. In the next section, we shall address the connection between \textbf{Felzenszwalb} and \textbf{LIME}.

\section{LIME}
\label{sec:lime}

\textbf{LIME} was primarily developed to address the "trusting a prediction" problem, which is vital for decision-making \cite{Tulio2016}. This is achieved by providing individual model prediction explanations: In terms of image classification, presenting visual artefacts that qualitatively correlate typical patches (in our case, \textbf{super-pixel}) of an image and its model prediction \cite{Tulio2016}.  

Before talking about \textbf{LIME} mechanism under the image classification scheme, we firstly review the definition of the following variables/functions \cite{Tulio2016}:

\begin{itemize}
    \item classifier f: a well-trained complex classification model (i.e., CNN)
    \item x ($x\in R^d$): original representations of an image instance awaited for model explanation.
    \item x' ($x' \in \{0,1\}^{d'}$): the binary vector for the interpretable representation of x, representing the "presence" (1) or "absence" (0) of image super-pixels formed via image segmentation (\textbf{Felzenszwalb} in this work).
    \item g ($g\in G$): an explanation as a model, where G is a family of interpretable models (i.e., linear models)
    \item $\Omega(g)$: model complexity of g (i.e., number of non-zero weights if g is a linear model).
    \item f(x): the probability that x belongs to a typical class.
    \item $\pi_x(z)$: a proximity measure between instance x and z.
    \item $L(f,g,\pi_x)$: a measure to quantify the ability of g to approximate f in the $\pi_x$ defined locality. The smaller the $L$, the better g has performed. 
    \item $z'$ ($z'\in (0,1)^{d'}$): a perturbed sample containing a fraction of the non-zero elements in $x'$.
\end{itemize}

To ensure g is both interpretable to human (low $\Omega(g)$) and faithful (low $L(f,g,\pi_x)$), \textbf{LIME} produces its model explanation by \cite{Felzenszwalb2004}: 

\begin{equation}
\label{eqn:lime_optimization}
\xi(x) =\mathop{argmin}\limits_{g\in G}\,\,\,\, L(f,g,\pi _x)+\Omega(g)
\end{equation}

Since \textbf{LIME} aims to perform \textbf{model-agnostic} model explanation, it samples instances around $x'$ by randomly drawing (hiding) non-zero elements of $x'$ and gives $z'$. By recovering $z'$ (weighted by $\pi_x(z)$) in the original representation $z$, one shall obtain f(z). f(z) then can be seen as a label for explanation model g. These perturbed samples and their corresponding labels could then be used to optimize Equation~\ref{eqn:lime_optimization} and finally obtain \textbf{$\xi(x)$}, the model explanation for the instance original representation x. One can then know which super-pixel in an image has positively/negatively contributed to class prediction, and hence evaluate whether the model predicts as humans do qualitatively.

\section{Application to Radio Galaxy Classification}
\label{sec:application}

Though \textbf{LIME} could be a useful model explanation method, there is a loose restriction of this technique: the user are often (not always) required to know what to expect before explaining a model. That is to say, a user should be aware of which features in an image contribute to object classification. Luckily, the FR classification problem we consider here has reasonably well-defined features for each class. In this case, \textbf{LIME} can be used to investigate the following questions:

\begin{enumerate}
    \item Did the model predict the image class mainly according to the central targeted source emission regions?
    \item Did the model consider irrelevant emission regions when making predictions?
    \item Did the model predict source class in accordance with those field regions responsible for typical source class morphology just as radio astronomers do?
\end{enumerate}

\begin{figure*}
  \centering
  \includegraphics[width=\textwidth]{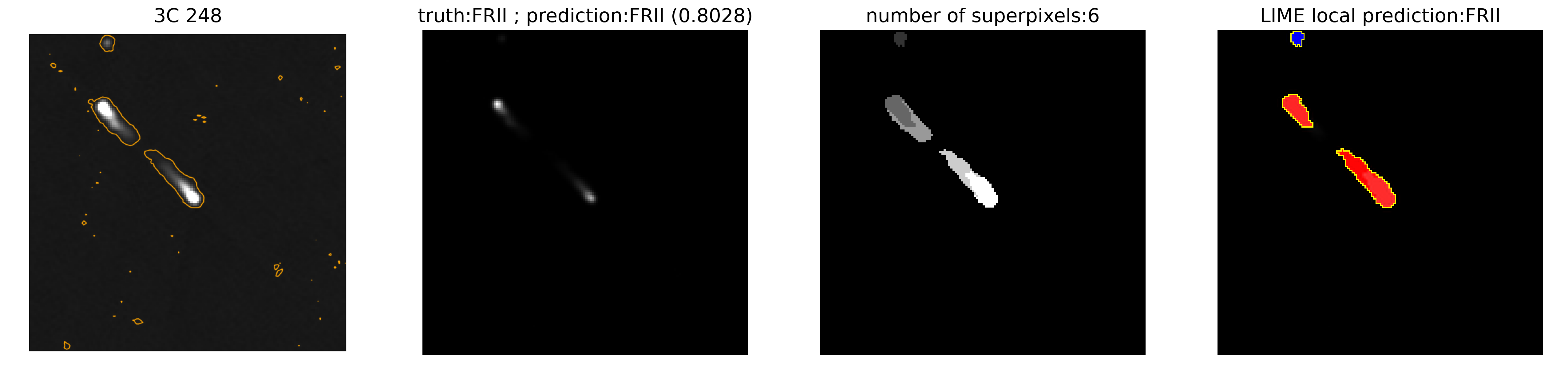}
  \caption{An illustration of sample image model prediction interpretation using \textbf{LIME}. From left to right: (1) The first picture shows the contour lines of the original FIRST radio survey image of 3C 248 at 3$\sigma$ level; (2) the second one shows the same image experinced normalization and was used in the model testing (downloaded from FR-DEEP v2); (3) the third picture illustrates the super-pixels generated by \textbf{Felzenszwalb} segmentation method, where the different color of the regions represent different super-pixels; (4) the last one is the interpretation of the picture using LIME. The red regions contribute positively to the FR\,II class prediction, while the blue one contributes negatively to the same prediction.}
  \label{fig:lime_example}
\end{figure*}

Figure~\ref{fig:lime_example} can be seen as an early example of the \textbf{LIME} model explanation in our work. It can be seen from the $4_{th}$ subplot of the figure that for the particular CNN classification algorithm in this work, the network has correctly identified 3C 248 as a FR\,II source, with both of its radio lobes contributing to FR\,II class prediction. On the top left of the plot, however, another separate source has contributed to the FR\,II classification negatively. In other words, when there is more than one radio galaxy in an image, the network can no longer claim "The central image source is a FR\,II radio galaxy". For our network, model user then should visual inspect these images with the aid of generated \textbf{LIME} maps. 

In terms of early statistical analysis, by visual inspecting the 130 samples in our data testset, we found the network does able to make prediction based on the image central object in most time, especially when an image contains one source only. It generally favors hot spots and those source radio lobes with relatively sharp margins when classifying a source as a FR\,II object, whereas the situation of FR\,I source classification is more complicated. \textbf{LIME} may also facilitate image mis-classification, though such diagnostics does not always succeed and thus require further investigations. Detailed discussions upon in what aspects can \textbf{LIME} help interpret our network predictions will be presented in Tang and Yue et al. (2023, in prep.).

\section{Conclusion}
\label{sec:conclusion}

We propose the use of \textbf{LIME}, a model-agnostic machine learning model interpretation technique to explain deep learning algorithm developed for Fanaroff and Riley radio galaxy morphology classification task. We present a routine of using this technique to explain model prediction behaviour of a trained CNN based classification algorithm. In this work, \textbf{LIME} generally segments image into multiple "super-pixels" via \textbf{Felzenszwalb} segmentation method, and find those super-pixels in an image that contributed to its model predicted image class. Our early analysis show that for our trained network:

\begin{itemize}
    \item predict image class mostly based on central source emission regions
    \item when more than one source presented in the same image, model prediction may be biased.
    \item FR\,II source classifications given by the network generally favor hot spots and source radio lobes with sharp margins.
\end{itemize}

Situations of FR\,I classification and mis-classification diagnostics are rather complicated, which require further investigation.

\section*{Acknowledgements}

HT gratefully acknowledges the support from the Shuimu Tsinghua Scholar Program of Tsinghua University; the fellowship of China Postdoctoral Science Foundation 2022M721875; and long lasting support of DoA TAGLAB research group, Tsinghua University and JBCA machine learning group. SY would like to acknowledge the teammate, for their wonderful collaboration and patient support; gratefully thanks  Hongsen Yin and Jianxiong Li for the helpful discussions and unreserved supports during the study. SY and ZW are also grateful to the cultivation of Strengthening Foundation Plan conducted by School of Physics and Astronomy, Sun Yat-sen University.

\end{document}